\documentclass[a4paper,titlepage]{article}

\usepackage[svgnames]{xcolor}
\usepackage{footnote}  
\usepackage[osf]{libertine} 
\usepackage[shortlabels]{enumitem}
\usepackage[numbers,sort&compress]{natbib}
\usepackage{arxiv}
\usepackage[utf8]{inputenc} 
\usepackage[T1]{fontenc}    
\usepackage{hyperref}       
\usepackage{url}            
\usepackage{booktabs}       
\usepackage{amsfonts}       
\usepackage{nicefrac}       
\usepackage{microtype}      
\usepackage{listings}
\usepackage{graphicx}
\usepackage{subcaption}
\usepackage{color}
\usepackage{float}
\usepackage[title]{appendix}
\usepackage{csquotes}

\newcommand{\code}[1]{\mbox{\lstinline{#1}}}
\title{An Efficient Implementation of Manacher's Algorithm}

\author{Shoupu Wan \\
\texttt{wanshoupu@gmail.com} \\
}

\begin{document}
    \maketitle

    \begin{abstract}
        Manacher's algorithm has been shown to be optimal to the longest palindromic substring
        problem.
        Many of the existing implementations of this algorithm, however, unanimously required
        in-memory construction of an augmented string that is twice as long as the original string.
        Although it has found widespread use, we found that this preprocessing is
        neither economic nor necessary.
        We present a more efficient implementation of Manacher's algorithm based
        on index mapping that makes the string augmentation process obsolete.
    \end{abstract}


    \section{Introduction}\label{sec:introduction}

    Finding a longest palindrome substring (LPS) in a given string is a fundamentally important
    question as it has widespread applications in mathematics, physics, chemistry, genetics, music,
    \textit{etc.}~\cite{Manacher:1975, shiu2010data, jeuring1993theories, galil1981string,
    crochemore1994text}.
    To one who is familiar with the song ``Rhythm of the Rain'', the prelude music might be
    very impressive.
    That is an example of musical palindromes.
    In genetics, palindromic sequences has an important capability---forming
    hairpins~\cite{chaos2008palindromelifeorigin}.
    It is amazing to learn that palindromes had played such an important role in life from the
    very beginning.
    But here I pick the LPS problem for two reasons.
    First, this problem is closely related to the study of symmetry.
    Often times, uncovering the underlying symmetry is the key for great solutions.
    This problem exemplifies how conscious application of
    mathematical analysis can help devise an algorithm.
    Secondly, this problem is a perfect case of study for demonstrating how to refactor messy and
    monolithic code with bloating duplications into a succinct and modular solution free
    from duplications step-by-step.
    Most of the techniques are discussed in depth in book~\cite{Wan:book}.

    Here is the structure of this article.
    In ~\autoref{sec:lps-statement}, we will set the problem statement.
    In ~\autoref{sec:reflection-symmetry}, the reflection symmetry with necessary
    mathematical context will be explained with aim at the application toward the LPS problem.
    In ~\autoref{sec:manacherAlgorithm}, Manacher's algorithm is presented together with some
    intuition.
    Then ~\autoref{sec:implementation-lps-augmented} we will discuss existing solutions
    with the string-augmentation preprocessing.
    In ~\autoref{sec:virtualMapping} and ~\autoref{sec:palindrom.modularized},
    we will present the new approach of index mapping to implement Manacher's algorithm.
    Also in these sections, we will perform multi-stage refactor process that eventually leads to
    a modulalr solution to the LPS problem with high readability.
    Finally in ~\autoref{sec:experiment}, we will put all the implementations presented in this
    article to test.
    We will compare the performance test result for different approaches.
    All solutions provided in this article will be implemented in Java.


\section{The problem Statement}\label{sec:lps-statement}

The LPS problem takes various forms in the literature.
For the sake of this article, we state the problem as
\begin{quote}
    ``Given an input string, find the longest palindromic substring in it (or one of them if
    there are more).''
\end{quote}
According to Merriam-Webster dictionary, a palindrome is ``a word, phrase, or sequence that reads
the same backward as forward''.
The length of a palindromic string can be either odd or even.
Accordingly, we may classify palindromic strings as odd or even.
For an odd palindrome, its center of symmetry, \textit{e.g.}, \emph{palindromic center}
or simply \emph{center}, falls on a character.
For a nonempty even palindrome, its center falls between two characters, which in this book
will be referred to as \emph{left center} and \emph{right center}, respectively.
Obviously, an emtpy string is also palindromic---it is the trivial case.
A palindromic substring (PSS) of a string is any substring that is a palindrome.
For a string of length $N$, there are $(2 * N + 1)$ palindromic centers, albeit some of
them may be trivial.
The sole PSS of an empty string is trivial.
The first and last PSS's of an nonempty string are trivial.
Apparently for a specific non-trivial palindromic center, there may be a series of co-centered
palindromic substrings.
We call the longest among these co-centered palindromic substrings
`\textit{prime palindromic substring}'.
Without loss of generality, we will limit our discussion on prime palindromic substrings only.

To some, solving the problem is not ``hard'' so to speak if optimality is not concerned.
One possible solution, for example, may be that
\begin{quotation}
    Iterate through each possible center and for each center, calculate the length of PSS\@.
    To calculate the length of PSS at a specific center, one can dispatch two indexes off the center
    outwardly in opposite directions symmetrically.
    If, at any step, a mismatch is encountered, stop.
    The substring lies between the two indexes.
\end{quotation}
The runtime complexity of such a naive solution is $O(N^2)$.
The difficulty about this problem is how to beat the quadratic runtime.
In his paper of 1975, Glenn Manacher discovered an algorithm with linear runtime.
It was later found that his method works not only for prefix PSS but for all PSS's.
This algorithm is now the so-called \textit{Manacher's algorithm}~\cite{Manacher:1975}.
We will dedicate the next few sections to get a thorough understanding for this algorithm.
First, we need a little bit of math about symmetry.

\section{Reflection symmetry}\label{sec:reflection-symmetry}

Reflection symmetry, \textit{a.k.a.}, mirror-image symmetry
\footnote{\url{https://en.wikipedia.org/wiki/Reflection_symmetry}} refers to spacial invariance
under a reflection transformation.
A reflection transformation is the operation that transforms coordinates to their
mirror-image \textit{w.r.t.} a fixed point, which we will refer to as the \emph{axis of symmetry} or
\emph{center of symmetry}.
In one-dimensional space ($1$D), a coordinate, $x$, and its transformed coordinate, $x^\prime$
\textit{w.r.t.} axis $a$ are related by equation:
\begin{equation}
    \label{eq:reflection-transform}
    x^\prime = 2a - x.
\end{equation}

\begin{figure}[H]
    \centering
    \caption{Two nearby axes of reflection symmetry partition
    the entire space periodically and infinitely.
    Overlapping palindromes also form similar periodic patterns.}\label{fig:ReflectionSymmetry}
    \begin{subfigure}{.4\textwidth}
        \includegraphics[width=\textwidth]{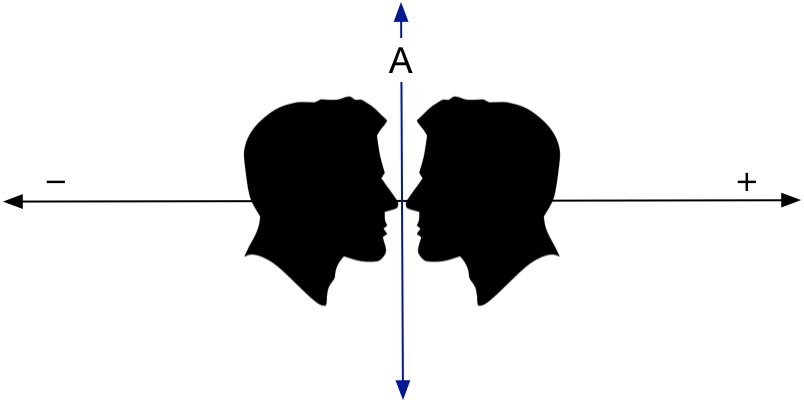}
        \caption{An axis of reflection symmetry}
        \label{fig:a-axis-reflection-symmetry}
    \end{subfigure}
    ~
    \begin{subfigure}{.4\textwidth}
        \includegraphics[width=\textwidth]{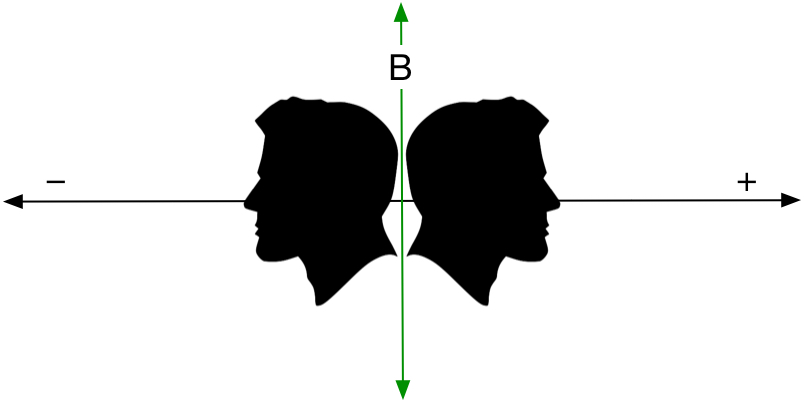}
        \caption{Another axis of symmetry}
        \label{fig:b-axis-reflection-symmetry}
    \end{subfigure}
    \vspace{1mm}
    \begin{subfigure}[b]{.4\textwidth}
        \includegraphics[width=\textwidth]{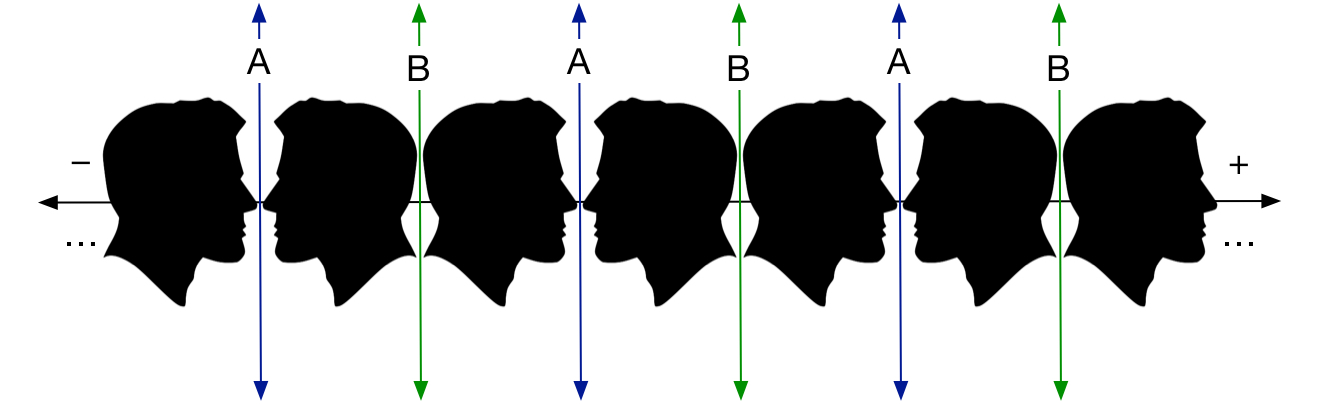}
        \caption{Two concurrent axes}
        \label{fig:two-axes-reflection-symmetry}
    \end{subfigure}
    ~
    \begin{subfigure}[b]{.4\textwidth}
        \includegraphics[width=\textwidth]{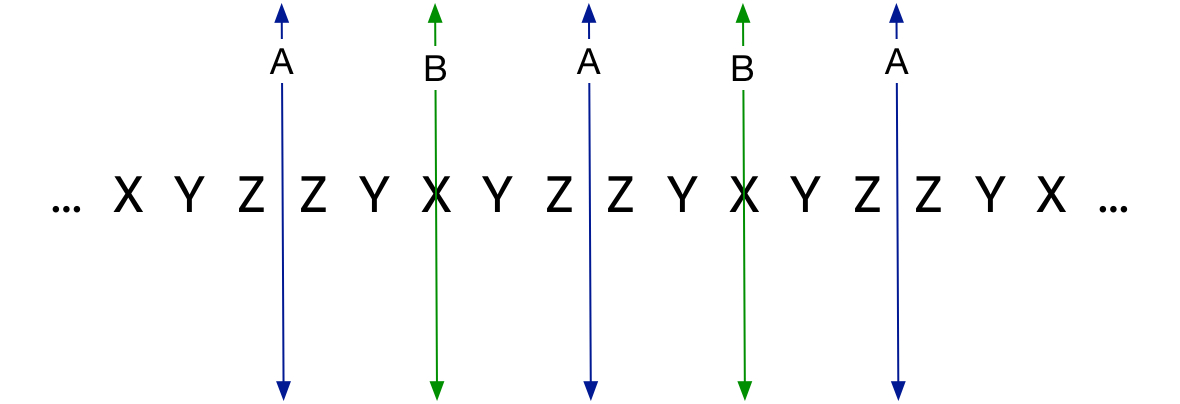}
        \caption{Overlapping palindromes}
        \label{fig:palindrome-symmetry}
    \end{subfigure}
\end{figure}

As shown in ~\autoref{fig:a-axis-reflection-symmetry}, axis of symmetry $A$ partitions the entire $1$D
space into $2$ half-spaces about itself---one on the left and one on the right.
There is a one-to-one mapping between the points in the $2$ half-spaces.
Axis $B$ does similarly (\autoref{fig:b-axis-reflection-symmetry}).
It is truly magical when two axes of reflection symmetry are present near each other
along the $x$-axis.
By repeatedly applying the reflection transformation, one may find infinitely many axes of symmetry
alternately along the $x$-axis, and collectively they partition the entire space
into periodic regions with period $2d$, where $d$ is the distance between the two axes of
symmetry (see~\autoref{fig:two-axes-reflection-symmetry}).
This effect may not be unfamiliar to
you if you have ever stepped in between two parallel mirrors---an array of clones of `you' appear,
alternately facing toward and away from you, aligned and coordinated.
This symmetric configuration in a discretized $1$D space resulting from multiple reflections
is the key intuition to the LPS problem.

\begin{figure}[tbh]
    \centering
    \caption{\label{fig:bananas-palindrome}
    Some examples of palindromic substrings in the string ``bananas'',
    labeled as `$1$`, `$2$`, and `$3$`.}
    \includegraphics[width=.5\textwidth]{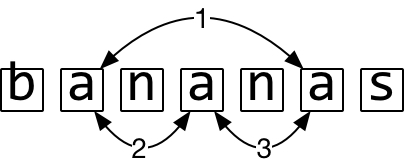}
\end{figure}

~\autoref{fig:palindrome-symmetry} shows an infinite palindromic string.
In reality, however, the aforementioned symmetry does not exist, as there is
no infinite space or string.
Nevertheless, the argument still holds for the finite string in the overlapping regions.
Of prime interest to us are a collection of `crowded', overlapping PSS's.
Under the wings of some large PSS's, some shorter palindromes may take shelter.
On the other hand, the larger PSS that cloaks over the shorter ones will project the latter's
mirror image to the opposite wing, because of reflection symmetry
(see ~\autoref{fig:bananas-palindrome}).
This reflective projection may be applied recursively as many times as there are
enclosing palindromes.
As a result, a substring $S$ may be projected to its mirror image $S^\prime$, which, in turn,
may be projected to its image $S^{\prime\prime}$ and so on.

\section{Manacher's Algorithm}\label{sec:manacherAlgorithm}

Equipped with this understanding of the reflection symmetry, we are in a better position to
crack the mystery of Manacher's algorithm.
Manacher's algorithm leans heavily on cached PSS's.
The reflective projection relates the to-be-calculated palindromic center with its mirror image in
the cache and this is the key step to avoid repeated character comparisons.

Taking string ``bananas'' for example as shown in ~\autoref{fig:bananas-palindrome},
knowing that PSS-$2$ has length $3$ and that PSS-$1$ mirrors (part of) PSS-$3$ to (part of)
PSS-$2$, we can skip all but the outermost pairs, which are `n' and `s'.
Upon seeing that they do not match, the length of PSS-$3$ is finally pinned at $3$.
So with one additional character comparison, we obtained the length of PSS-$3$.
That is where savings come from.

To sum up, let us iterate through the string from left to right and cache the result in an array,
\textit{e.g.},
\begin{center}
    \begin{tabular}{cccccccccccccccc}
        index:&0,&1,&2,&3,&4,&5,&6,&7,&8,&9,&10,&11,&12,&13,&14\\
        length:&0,&1,&0,&1,&0,&3,&0,&5,&0,&3,&0,&1,&0,&1,&0
    \end{tabular}
\end{center}
At each step, we also keep a reference PSS which reaches the farthest right.
When examining a new center, we first check its mirror image with respect to the
reference.
Depending on the relationship between its mirror image PSS and the reference PSS,
some or all character comparisons for the new center may be spared, just as in the case of
``bananas''.
With no more ado, lists Manacher's algorithm.
\begin{enumerate}
    \item Initialize an array \code{pss} of size \code{2 * n + 1} and element at index \code{i}
    stores length of $\mbox{PSS}(i)$.
    \item Initialize \code{refCenter = 0}, which stores the palindrome center whose right wing
    reaches rightmost in each iteration of the main loop;
    \item For each character at \code{j = 0, 1, ..., n} in the augmented array,
    \begin{enumerate}[label={\roman*:}]
        \item If \code{j} lies outside of \code{pss(refCenter)} to the right,
        calculate $\mbox{PSS}(j)$ from scratch;
        update \code{refCenter} accordingly;
        skip to the next iteration.
        \item Otherwise, find the mirror image, \code{k}, of \code{j}
        \textit{w.r.t.} \code{refCenter}.
        \item If $\mbox{PSS}(k)$ is completely contained in \mbox{PSS(\code{refCenter})},
        then \code{pss(k)=pss(j)};
        \item Otherwise, we need to calculate the $\mbox{PSS}(j)$, but only the portion
        outside of \mbox{PSS(\code{refCenter})}, if any.
    \end{enumerate}
\end{enumerate}

\section{Augmented-String Implementations}\label{sec:implementation-lps-augmented}

I hope you have already grasped the gist of Manacher's algorithm before we talk about its
implementations.
Traditionally the implementations of Manacher's algorithm assumes augmenting the original string
by inserting a dummy character between each adjacent pair of characters in the
original string.
For uniformity, we also add dummy characters at the ends (insert one dummy
character in the case of an empty string).
By doing so, we established a one-to-one mapping between the PSS's in the original string
and the odd PSS's in the augmented string.
So for \mbox{``bananas''}, the augmented string would be \mbox{`` b a n a n a s ''}
if blank space is chosen as the augmenting character.
It is important that the chosen dummy character be absent from the original string.
Otherwise, spurious result may result.

With the literally augmented string, implementing Manacher's algorithm becomes straightforward.
One can refer to several published implementations in different programming languages.
There is a Java version by the CS department of Princeton
University\footnote{\url{https://algs4.cs.princeton.edu/53substring/Manacher.java.html}}.
There is a Python version by Fred
Akalin\footnote{\url{https://www.akalin.com/longest-palindrome-linear-time}}.
There is even a Haskell implementation along with a discussion of the algorithm itself
in Johan Jeuring's blog\footnote{
\url{http://finding-palindromes.blogspot.com/2012/05/finding-palindromes-efficiently.html}} and
also book~\cite{jeuring1993theories}.
Lastly, an implementation of my own is also provided for the sake of reference
(~\autoref{lst:LongestPalindromeAugment}).

\begin{lstlisting}[label={lst:LongestPalindromeAugment},language=Java,escapechar=~,
caption={An implementation of Manacher's algorithm based on literally augmented string.}]
public String longestPalindrome(String input) {
    final int n = input.length() * 2 + 1;
    char[] aug = new char[n];
    for (int i = 0; i < n; ++i) {
        aug[i] = (i & 1) == 0 ? 0 : input.charAt(i / 2);
    }
    int maxStart = 0, maxEnd = 0;
    int[] pss = new int[n];
    pss[0] = 1;
    for (int center = 0, j = 1; j < n; ++j) {
        int wing = pss[center] / 2;
        if (j < center + wing) {
            int jmirror = 2 * center - j;
            int jwing = pss[jmirror] / 2;
            if (jmirror - jwing > center - wing) {
                pss[j] = pss[jmirror];
                continue;
            }
        }
        //get the length of the LPS centered on j
        for (int i = center + wing + 1; ; i++) {
            if (2 * j - i < 0 || i == n || aug[2 * j - i] != aug[i]) {
                pss[j] = 2 * (i - 1 - j) + 1;
                break;
            }
        }
        center = j;
        if (maxEnd - maxStart < pss[center] / 2) {
            maxStart = (center + 1 - pss[center] / 2) / 2;
            maxEnd = (center + 1 + pss[center] / 2) / 2;
        }
    } //end for
    return input.substring(maxStart, maxEnd);
}
\end{lstlisting}

\section{Virtual Augmentation}\label{sec:virtualMapping}

Even though the string-augmentation approach has found widespread use for
the implementation of Manacher's algorithm, this is neither convenient nor necessary.
For large strings such as DNA chains in genome sequencing, it is costly to have to
construct the augmented string with doubled memory footprint~\cite{shiu2010data}.
Furthermore, it is onerous and sometimes quite annoying to have to identify a suitable dummy
character for the augmentation process.
In this section, we seek a more concise and economic way to
implement Manacher's algorithm---sparing the string augmentation process.

\begin{savenotes}
    \begin{table}[htb]
        \def\arraystretch{1.5}
        \centering
        \caption{\label{tab:semantics-arithmetic} Semantics of index arithmetic expressions given
        \code{i} being the palindromic center and \code{x} an arbitrary index, both in the augmented
        string; \code{pss} is the array storing lengths of palindromic substrings.}
        \begin{tabular}{r|p{.5\textwidth}|l}
            \hline
            Arithmetic & Semantic & Helper Function \\
            \hline
            \code{i / 2} & The left center in the original string (only for even \code{i}) \\
            \code{(i - 1) / 2} & The center in the original string (only for odd \code{i}) \\
            \code{2 * i - x} & Mirror image of \code{x} about the center in the augmented string
            & \code{toMirrorImage} \\
            \code{i - pss[i]} & Left bounding index in the augmented string
            & \code{getLeftBound} \\
            \code{i + pss[i]} & Right bounding index in the augmented string
            & \code{getRightBound} \\
            \code{(i - pss[i]) / 2} & Left bounding index in the original string\footnote{
            \label{odd-odd-even-even} This is because if \code{i} is even, length of the
            PSS centered on \code{i} can only be even.
            Conversely, if \code{i} is odd, length of the PSS centered on \code{i} can only be
            odd.} \\
            \code{(i + pss[i]) / 2} & Right bounding index in the original string
            \textsuperscript{\ref{odd-odd-even-even}} \\
            \hline
        \end{tabular}
    \end{table}
\end{savenotes}

The key is to establish an index mapping between the original and the augmented string.
Consider the string \mbox{``bananas''} (~\autoref{fig:bananas-palindrome}) and its augmented string
\mbox{`` b a n a n a s ''}.
Let us try to establish the correspondence rules between the original string and the augmented
string.
First, each character in the original string corresponds to a character in the
augmented string with an odd index.
So there is a one-to-one correspondence between the
indexes of the original string and the odd indexes of the augmented string.
The even indexes of the augmented string, however, corresponds to the inner boundaries between
adjacent characters.
Together, they establish a one-to-one mapping between PSS's of the original string and odd
PSS's in the augmented string.
It is easy to see that a PSS in the
augmented string are completely determined by the corresponding PSS in the original string.
The added augmentation characters do not interfere at all.
Therefore, if we can formulate the PSS's of the original string in terms of the indexes of
the hypothetically augmented string, we would be freed from the need to construct the augmented
string in memory.
We name this method ``index mapping''.
Accordingly, the process of relying on mapped indexes for calculating
palindromic substrings are named ``virtualized augmentation''.
Not only does virtualized augmentation makes the double-sized memory consumption obsolete,
but it also frees us from the burden of choosing dummy characters.
Some arithmetic expressions and their semantic meanings have been tabulated in
~\autoref{tab:semantics-arithmetic} for reference.
Based on the idea of virtual augmentation, we come up with a new approach to implement
Manacher's algorithm (~\autoref{lst:LongestPalindrome}).
Note that the solution listed in ~\autoref{lst:LongestPalindrome} has
$O(N)$ runtime as well as memory complexity.

\begin{lstlisting}[label={lst:LongestPalindrome},language=Java,escapechar=~,
caption={Implementation of Manacher's aglorithm using index mapping.}]
public String longestPalindrome(String input) {
    int[] pss = new int[input.length() * 2 + 1];
    for (int i = 1, refCenter = 0; i < pss.length; ++i) {
        // refCenter = center reaching to the right farthest
        if (refCenter + pss[refCenter] <= i) {
            pss[i] = palength(input, i, i);
            refCenter = i;
        } else {
            int im = refCenter * 2 - i;
            //assert im >= 0
            if (im - pss[im] > refCenter - pss[refCenter]) {~\label{line:lps-soft-duplication}~
                pss[i] = pss[im];
            } else {
                pss[i] = palength(input, i, refCenter + pss[refCenter]);
                refCenter = i;
            }
        }
    }

    for (int i = 0, refCenter = 0; ; ++i) {
        if (i == pss.length) {
            return substring(input, refCenter, pss[refCenter]);
        }
        if (pss[i] > pss[refCenter]) {
            refCenter = i;
        }
    }
}

String substring(String str, int c, int len) {
    //if c is even, len can only be even
    //if c is odd, len can only be odd
    int s = (c - len) / 2;
    int e = (c + len) / 2;
    return str.substring(s, e);
}

int palength(String s, int c, int i) {
    final int n = 2 * s.length() + 1;
    for (int mi = 2 * c - i; ; ++i, --mi) {
        if (mi < 0 || i >= n || (i & 1) == 1 && s.charAt(i / 2) != s.charAt(mi / 2)) {
            return i - c - 1;
        }
    }
}
\end{lstlisting}

\section{Solutions for Readability}\label{sec:palindrom.modularized}

By virtual augmentation, we resolved the memory footprint issue and the shadowy dummy character.
The mission is well accomplished.
But by no means should we settle here.
The code in ~\autoref{lst:LongestPalindrome} is like a bowl of spaghetti noodle, isn't it?
Even though the code is divided up into three functions, its cleanliness still suffers.
It takes some courage for me to read it and try to figure out what each line does in just a couple
days after I wrote it.
I can not imagine it would be easier for one who has just come across the code.
A principle that all developers should stick to is
\enquote{If it is not readable, it is not acceptable}.
Our next goal is to seek a more readable way to implement it.

At a glance, the code is packed with arithmetic expressions, some for symmetry
transformations, some for key look ups, \textit{etc.}
These are all resulted from the virtual augmentation.
But if you look carefully, you may spot bloating repetitions of some arithmetic operations.
Some are hard ``copy-n-paste'' of others while
more belong to the category of so-called ``soft duplicate''~\cite{Wan:book}.
Another problem with the implementation is that it is monolithic.
The same function does too many things at once, violating the Single Responsibility
Principle~\footnote{\url{https://en.wikipedia.org/wiki/Single-responsibility_principle}}.
A well organized solution in this context should consist of a group of meaningful, single-purposed,
and reusable functions.

So we have two tasks that are somehow related---one for elimination of duplicate code and the
other to make the solution more modular.
Our starting point is ~\autoref{lst:LongestPalindrome}.
We may approach both tasks first with an understanding of the semantics of some operations,
especially those that are repeated.
If it helps, we can factor them out into helper functions.

Take as an example line ~\autoref{line:lps-soft-duplication} in ~\autoref{lst:LongestPalindrome}:
\begin{lstlisting}
    if (im - pss[im] > refCenter - pss[refCenter])
\end{lstlisting}
It may be obscure to untrained eyes.
But the pattern is ``Given an index, get the result as the index minus the element
at the index''.
So we can factor that out into a function, \textit{e.g.}, \code{getLeftBound}
(see~\autoref{tab:semantics-arithmetic}).
All occurrences of the same logic may be replaced by a call of function \code{getLeftBound}.
This alone helps get rid of $3$x duplications.
In fact, similar refactor may be performed for other entries in
~\autoref{tab:semantics-arithmetic}.
By doing so, we first modularized the solution by creating succinct, easily understandable helper
functions.
The helper functions, in turn, may be reused to reduce code duplication.
One stone for two birds.
The technique is discussed in length in the book~\cite{Wan:book}.

\begin{lstlisting}[label={lst:lpsModular-main},firstline=3,language=Java,escapechar=~,
caption={Modularized solution of the LPS problem.}]
public class LongestPalindromeSolver {

    private String input;
    // The i-th element in array 'lsp' records the max length of palindrome centered
    // on a char or between two adjacent chars of the input string depends on the parity of i:
    // if i is even, it is centered between the (i/2)th char and the (i/2 + 1)th char;
    // if i is odd, it is centered on the (i-1)/2th char in string input
    // pss[0] = 0 by definition
    private int[] pss;

    public String longestPalindrome(String input) {
        this.input = input;
        pss = new int[input.length() * 2 + 1];
        solve();
        return substring(argmax());
    }
}
\end{lstlisting}

Our refactored solution---the \code{class LongestPalindromeSolver}---is listed in
~\autoref{lst:lpsModular-main} and ~\autoref{lst:lpsModular-helper}.
The class has two private fields, one for input and the other output.
The only point of entry is the public method \code{longestPalindrome} which
helps with bookkeeping and dispatching.
All others are helper functions listed in ~\autoref{lst:lpsModular-helper} and
explained below.

\begin{lstlisting}[label={lst:lpsModular-helper},language=Java,escapechar=~,
caption={The refactored helper functions.}]
/*
 * Implement Manacher's algorithm
 */
void solve() {
    for (int i = 1, refCenter = 0; i < pss.length; ++i) {
        // refCenter = center reaching to the right farthest
        if (getRightBound(refCenter) <= i) {
            pss[i] = palength(i, i);
            refCenter = i; // i becomes the rightmost palindrome
            continue;
        }
        int mi = toMirrorImage(refCenter, i);
        //assert mi >= 0
        if (getLeftBound(mi) > getLeftBound(refCenter)) {
            // palindrome is wrapped
            pss[i] = pss[mi];
        } else {
            //calculate the part outside
            pss[i] = palength(i, getRightBound(refCenter));
            refCenter = i;
        }
    }
}

int getLeftBound(int i) {
    return i - pss[i];
}

int getRightBound(int i) {
    return i + pss[i];
}

int toMirrorImage(int axis, int x) {
    return 2 * axis - x;
}

/*
 * Determine if there is a mismatch between virtual char at i and j
 * A mismatch happens if both indexes are even and the charAt not equal
 */
boolean isMismatch(int i, int j) {
    return (i & 1) == 1 && input.charAt(i / 2) != input.charAt(j / 2);
}

/*
 * Find the palindrome in string input centered at center.
 */
int palength(int center, int index) {
    for (int mi = toMirrorImage(center, index); ; ++index, --mi) {
        // if one of the conditions: reaching the left end,
        // reaching the right end, or encountered char mismatch
        if (mi < 0 || index >= pss.length || isMismatch(index, mi)) {
            return index - center - 1;
        }
    }
}

String substring(int center) {
    int left = getLeftBound(center) / 2;
    int right = getRightBound(center) / 2;
    return input.substring(left, right);
}

/*
 * Find the index of maximum palindromic substring
 */
int argmax() {
    int maxi = 0;
    for (int i = 0; i < pss.length; ++i) {
        if (pss[i] > pss[maxi]) {
            maxi = i;
        }
    }
    return maxi;
}
\end{lstlisting}

The \code{solve} function implements the large part of Manacher's algorithm.
Methods \code{getLeftBound}, \code{getRightBound}, \code{palength}, and \code{toMirrorImage}
each corresponds with an index mapping expressions in \mbox{~\autoref{tab:semantics-arithmetic}}.
Method \code{isMismatch} checks for pairwise character mismatch.
Method \code{substring} helps construct the final result---the longest palindromic substring.
Method \code{argmax} is, as suggested by its name, a quick-and-dirty implementation of the $argmax$
mathematical function.

\section{Experiment}\label{sec:experiment}

To test the performance of the implementations and catch possible regressions,
we designed a simple experiment to compare the implementations listed in this article.
We randomly generated strings of various lengths $L$ using $A$ alphabets as the testing benchmarks
for $L = 1K, 10K, 100K, 1M, 10M, 100M, 1B$ and $A = 2, 3, 5, 8, 13, 21$.
To reduce error, each run is repeated three times and the average is taken.
We found no strong correlation between runtime and
$A$ or the length of longest palindromic substrings.
The linearity of the runtime vs size of input string stands out quite obviously
in ~\autoref{tab:perf-test} which is expected.
In summary, our new index-mapping based implementations perform similarly as the approach based on
string-augmentation but is more efficient in terms of memory footprint.
Where the implementation with augmented string fails, the virtualized augmentation approach still
runs successfully.
Besides the readability, there is also slight improvement in runtime in our modularized solution.

\begin{table}[H]
    \def\arraystretch{1.5}
    \centering
    \caption{\label{tab:perf-test} Runtime measurement for
    the three implementations ``String Augmentation'' ~\autoref{lst:LongestPalindromeAugment},
    ``Index Mapping'' ~\autoref{lst:LongestPalindrome}, and ``Modularized Index
    Mapping''~\autoref{lst:lpsModular-main}.}
    \begin{tabular}{r|c|c|l}
        \hline
        Length & String Augmentation & Index Mapping & Modular Solution \\
        \hline
        $1000$ & $0.10$ & $0.09$ & $0.09$ \\
        $10000$ & $0.10$ & $0.09$ & $0.09$ \\
        $100000$ & $0.11$ & $0.12$ & $0.12$ \\
        $1000000$ & $0.17$ & $0.18$ & $0.18$ \\
        $10000000$ & $0.43$ & $0.44$ & $0.42$ \\
        $100000000$ & $3.29$ & $3.61$ & $3.24$ \\
        $1000000000$ &\code{OutOfMemory}& $36.86$ & $33.51$ \\
        \hline
    \end{tabular}
\end{table}

    \section{Conclusion}\label{sec:conclusion}

    In conclusion, we went through the longest palindromic substring
    problem as a case of study.
    We discussed the reflection symmetry required to understand Manacher's algorithm.
    We presented a novel implementation of Manacher's algorithm that avoided the tedious and
    costly string augmentation with index mapping.
    We compared the performance of the new approach against that of string-augmentation in
    terms of memory and runtime complexities.
    Using the techniques presented in previous chapters of this book, we refactored the monolithic
    solution with bloating duplication into a more modular and readable one.
    \bibliographystyle{unsrt}
    \bibliography{article}
\end{document}